\documentclass[a4,12pt]{article}
\usepackage{amssymb}
\usepackage{cite}
\def\be{\begin{eqnarray}}
\def\ee{\end{eqnarray}}
\def\bes{\begin{eqnarray*}}
\def\ees{\end{eqnarray*}}
\def\ba{\begin{array}}
\def\ea{\end{array}}
\def\beenu{\begin{enumerate}}
\def\eenu{\end{enumerate}}
\def\bt{\begin{tabular}}
\def\et{\end{tabular}}
\def\beitem{\begin{itemize}}
\def\eitem{\end{itemize}}

\def\vev{{\it vev}}

\def\tr{{\mbox{tr}\,}}

\def\rank{{\mbox{rank}\,}}
\def\diff#1#2{{\frac{\partial #1}{\partial #2 }}}

\def\BL{$(B-L)$}
\def\ot{$\otimes$}
\def\op{{$\oplus \, $}}
\def\flip{flipped $SU(5)\otimes U(1)$}
\def\GUT{$SO(10)$}

\newcommand{\GW}{$SU(3)\otimes SU(2) \otimes U(1)$}
\def\GG{$SU(5)$}
\def\Pl{\mbox{Planck}}
\def\latin#1{{\it #1}}

\def\ie{\latin{i.e.}}

\def\eg{\latin{e.g.}}
\def\adhoc{\latin{ad hoc}}

\def\irr#1{\underline{\bf #1}}

\begin{document}

\begin{titlepage}

\begin{flushright}
Saclay T02/025 \\
hep-th/0202278 \\
\end{flushright}

\vskip 1cm
\begin{center}
{\Large{\bf Intermediate Symmetries in the Spontaneous Breaking of
Supersymmetric $SO(10)$ }}
\end{center}
\vskip1.0cm

\vskip 24pt
\centerline{{\bf F. Buccella}}
\vskip 8pt
\centerline{Dipartimento di Scienze Fisiche dell'Universit\`a Federico II}
\centerline{and INFN, Sezione di Napoli, Naples, ITALY}
\vskip 12pt
\centerline{{\bf C.A. Savoy}}
\vskip 8pt
\centerline{CEA, Service de Physique Th\'eorique, CE-Saclay}
\centerline{F-91191 Gif-sur-Yvette Cedex, FRANCE}
\vglue 1.2truecm
\begin{abstract}
We study the supersymmetric spontaneous symmetry breaking of \GUT\
into \GW\  for the most physically interesting cases of \GG\  or \flip\
intermediate symmetries. The first case is more easily realized while
the second one requires a fine-tuning condition on the parameters of
the superpotential. This is because in the case of \GG\  symmetry there
is at most one singlet of the residual symmetry in each \GUT\
irreducible representation. We also point out on more  general grounds
in supersymmetric GUT's that some intermediate symmetries can be
exactly realized and others can only be approximated by fine-tuning. In
the first category, there could occur some tunneling between the vacua
with exact and approximate intermediate symmetry. The \flip\ symmetry
improves the unification of gauge couplings if \BL\ is broken by $\|
(B-L) \| =1$ scalars yielding right handed neutrino masses  below
$10^{14}\, {\rm GeV}$.

\end{abstract}
\vfill

\vfill
\footnoterule
\noindent
{}

\end{titlepage} 


The experimental data suggest at least two new high scales in particle
physics. On one hand, the interpretation of the solar \cite{sun} and
atmospheric \cite{atm} neutrino anomalies in terms of oscillations
\cite {pont} require (mass$)^2$ differences which can be accounted for
in the framework of the see-saw mechanism with very heavy right-handed
neutrinos.  Their Majorana masses settle a high scale, $M_R$, to be
associated with the violation of the lepton numbers. On the other hand,
the extrapolation of the three running gauge couplings of the Standard
Model, suggest that they converge towards a common value at a very high
scale, $M_U$, giving evidence for a grand-unifying symmetry. The
extraction of $M_R$ from the neutrino data suffers from uncertainties
\cite{abuc}, while $M_U$ depends on the physical states that are
assumed at intermediate energies to improve the three-to-one
convergence of the gauge couplings.  Still, they should not differ by
more than a few orders of magnitude, not so much as compared to the
huge hierarchy between these scales and the electroweak symmetry
breaking scale.  It is tempting to associate these two scales to the
spontaneous breaking of some very high energy gauge symmetries.

Despite the relative vicinity of the two scales, there is no compelling
reason to embed the gauge symmetries of \GW  into a larger one. In
particular, this is not necessary to explain the gauge coupling
unification in a string theoretical framework. Nevertheless,
grand-unification symmetries  are a very attractive hypothesis (as far
as one has control of the proton lifetime) with predictive power. The
natural GUT symmetry encompassing both  the Standard Model gauge group
\GW\   and a gauged $B-L$ symmetry is \GUT\ \cite{gut}. It goes without
saying,  this is not the only motivation for a \GUT\  GUT, and many
other aspects are to be found in the huge literature on this subject
\cite{moha}.

The study of the spontaneous breaking of the non-supersymmetric \GUT\
models with the present values of the strong coupling $\alpha _s$,
shows that the \BL\  symmetry has to be broken at an intermediate scale
around $10^{10}-10^{12}\, {\rm GeV}$ \cite{aprt} to allow for the
\GUT\  unification.  The consistency of this relatively low value with
neutrino mass patterns has also been discussed in this context
\cite{abuc}.

In this paper we present a reappraisal of these matters in the framework of
supersymmetric \GUT\  GUT's. We point out that with the same set of fields
that can produce the breaking of \GUT\ into \GW , there are other vacua with 
intermediate gauge symmetries, \eg , the Georgi-Glashow \GG . When the \GW\
vacuum approaches the \GG\ one it has an approximate Georgi-Glashow
symmetry, but if they get too close, the physical \GW\ vacuum would be
tunneled into the \GG\ one. Instead, there are other possible intermediate
symmetries, \eg , \flip , which can only be approximated by tuning the
parameters in the superpotential, so that they are not expected to be well
realized. An approximate \GG\ symmetry would correspond to the breaking of \BL\
above the gauge coupling unification scale, an approximate \flip\ symmetry to
the opposite situation. In either cases, a big difference in these scales would
conflict with the seesaw interpretation of the neutrino oscillation data. Hence
a control of either the tunneling or the tuning is needed. Fortunately, the gauge 
coupling unification points in the direction of a moderate difference in these 
scales, as the neutrino oscillations seem to do as well. 

It is well known that coupling  unification is -- almost -- realized by
the minimal supersymetrization of the Standard Model degrees of freedom
around $ 1\, {\rm TeV}$, consistently with a \GG\  unification.
Therefore, any intermediate symmetry between \GW\ and \GUT\  should
approximately preserve the SUSY \GG\  prediction for the gauge
couplings. Actually, an accurate evaluation of the gauge coupling
running at two-loops displays a strong model dependence on the
supersymmetric particle thresholds. For instance, a recent analysis
\cite{pok} shows that in the MSSM with universal gaugino and scalar
masses at the TeV scale, the exact two-loop coupling unification would
occur for $\mu \sim 10^{4}\, {\rm GeV}$, where $\mu$ is the usual MSSM
higgsino  mass parameter. For lower values, $\alpha_s ({\rm M_Z})$
comes out slightly  higher than the experimental data. The first point
we would like to make here is the possibility to improve the prediction
of $\alpha_s ({\rm M_Z})$  if one assumes that $SO(10)$ is broken into
the ``flipped'' $SU(5)\otimes U(1)$  symmetry which is then broken at a
slightly smaller scale \footnote{The \flip\ gauge symmetry has other
well-known appealing aspects \cite {enz},  including an elegant
mechanism to obtain the doublet-triplet splitting.}.

Although the present precision on $\alpha _s ({\rm M_Z})$ requires a
two-loop calculation, a one-loop study is sufficient for the
qualitative argument presented here. Let us first define the standard
combinations of $b$ parameters that control the approach to coupling
unification, namely, the running of $\alpha / \alpha _s$ and
$\sin{\theta_W}$, respectively:  $\Delta_s b = b_3 - 3b_2 /8 - 5b_1 /8$
and $\Delta _w b = b_2 - b_1$. In the Standard Model, the ratio $\Delta
_s b / \Delta _w b = 1.15$, so that, for non-supersymmetric
grandunification, one needs new physics at intermediate scales with a
larger value of $\Delta _s b / \Delta _w b .$ With the addition of the
supersymmetric partners within the MSSM, this ratio increases to
$1.34$. If the supersymmetry threshold is below 1 TeV, the two-loop
prediction for $\alpha _s ({\rm M_Z})$ turns out to be just a bit above
the experimental value. Hence, one can improve the gauge coupling
unification by adding new degrees of freedom at a scale just below
$M_U$, with a relatively low value of $\Delta _s b / \Delta _w b$.

The \flip\ model has $\Delta _s b / \Delta _{21} b = 5/8$, as $b_3 =
b_2 = b_5$.  Therefore, with the symmetry breaking pattern given by
\GUT\ $\rightarrow$ \flip\ at the scale $M_U$, and then
\flip\ $\rightarrow$ \GW , at the scale $M_R$, one can tune $\alpha _s
({\rm M_Z})$ toward its experimental value. The ratio $r = M_R / M_U$
depends on the effective supersymmetry threshold $T_{SUSY} $, that
incorporates the  various supersymmetric particle masses. Roughly
speaking one is correcting the supersymmetry threshold dependence with
the approximate \flip\ threshold effects and $r$ cannot be very small.

This symmetry breaking scheme can be implemented by introducing fields
in either a \irr{45} or a \irr{210} representation, for the first
breaking at $M_U$, and fields in either a \irr{16} \op \irr{16*} \op
\irr{54}, or a \irr{126} \op \irr{126* } \op \irr{54} representation,
for the second one, at $r M_U$. However, the \irr{126} \op \irr{126*},
which would yield a $\|\Delta (B-L) \| =2$ breaking, cannot be added,
since it would give $\Delta _s b < 0$.  It remains the only possibility
of a \irr{16} \op \irr{16*} breaking with $\|\Delta (B-L) \| =1 $. With
$r < 1$ this is the only simple pattern that can improve the gauge
coupling unification with a reasonable sparticle spectrum and that
would correspond to a breaking of $(B-L)$ slightly below the gauge
coupling unification scale. In particular, the option of left-right
symmetric sub-groups of $SO(10)$ would lead to gauge coupling
unification only if the sparticle masses would be above $10^4$GeV, a
situation requiring fine-tuning of the MSSM parameters to yield the
electroweak symmetry breaking scale.

With these motivations for our study of the supersymmetric spontaneous
breaking of \GUT\ into \GW\ with \GG\ or \flip\ intermediate
symmetries, let us first stress some general properties of the gauge
symmetry  breaking in supersymetric theories. The minimum conditions:
\be
W_i (z^0 ) = \diff{W(z)}{z^i}(z^0 ) = 0 ~ , \label{min}
\ee
where $W(z)$ is the superpotential and the $z^i$'s stand for the
components of the complex scalar fields, are non-trivial only for the
components  $W_i (z^0 )$ along the directions invariant under the
little group $H_{z^0}$ of $z^0$. This follows from the invariance of
$W(z)$ under $H_{z^0} .$ The gradient directions along all $H_{z^0}$
singlet fields must be considered.  Therefore, the number  $n$ of
non-trivial equations is equal to the number of $H_{z^0}$ singlets in
the representation of the chiral multiplets.  Generically, the
solutions of the resulting system of $n$ equations and $n$ variables
are linear combinations of all the singlet fields, in proportions fixed
by the parameters in the superpotential. (We concentrate here on gauge
symmetries, but this remark applies to the  -- complexified of the --
global symmetries of the superpotential as well.)

If the initial gauge group is $G_U$ and $H_{z^0}$ is $G_0$ ( we shall
consider later on the case where they are \GUT\ and \GW , respectively)
we may look for solutions with symmetry $G_I$, with $ G_0 \subset G_I
\subset G_U ,$ if there are $p < n$ $G_I$ singlets among the fields,
since in this case the number of non-trivial equations also reduces to
$p .$ We may also look for solutions whose exact symmetry is $G_0$
which possess an approximate symmetry $G_I$ because the predominant
\vev 's are the $G_I$ singlets. When these solutions approach the
corresponding one with exact $G_I$ symmetry, a tunneling between the
two vacua may become possible. Instead, a vacuum with an approximate
symmetry $G_I$ has not necessarily a counterpart with exact symmetry
$G_I$.

Let us illustrate these situations in a model with $G_U =$\GUT\  and $
G_0 =$\GW\  , with the Higgs chiral multiplets in a \irr{16} \op
\irr{16*} \op \irr{45} \op \irr{54} representation of \GUT\ ,
corresponding to the spinors $\psi$ and $\bar{\psi} ,$ the
antisymmetric matrix $A$ and the symmetric matrix $S,$ respectively.
Their \GW\ singlets are the following five directions:

\noindent {\it a)} $\psi_1 \in$ \irr{16} and $\bar{\psi_1} \in$
\irr{16*}, with little group \GG , which by definition is the
Georgi-Glashow one,

\noindent {\it b)} $A_1$ and $A_{24}$ in the \irr{45} transforming as a
singlet and a \irr{24} under this \GG , respectively,

\noindent {\it c)} $S_0 \in$ \irr{54} , with Pati-Salam little group
$SO(6)\otimes SO(4) = SU(4)\otimes SU(2)\otimes SU(2).$

The \irr{45} components $A_1$ and $A_{24}$ can be rearranged into a
singlet and a \irr{24} component with respect to the \flip\ as
follows:
\be 
A'_1=\frac{1}{5}A_1+\frac{2\sqrt{6}}{5} A_{24} \\ \nonumber
A'_{24}=\frac{2\sqrt{6}}{5} A_1 - \frac{1}{5} A_{24}
\label{45flip}
\ee
As a linear combination of $A_1$ and $A_{24} , $ $A'_1$ belongs to the
same critical orbit as $A_1$, \ie , they are related by a
\GUT\ rotation which does not leave $\psi_1$ invariant.

The most general superpotential with quadratic and cubic invariants
has the form:
\be
W= m \bar{\psi}\psi + M \tr A^2 + \mu\tr S^2 + h \bar{\psi} A\psi
+ \lambda \tr A^2 S + \kappa\tr S^3  \label{w}
\ee 
In Table 1 only the contributions of the \GW\  invariant fields to
these invariants are written.

The non-trivial equations obtained from the conditions (\ref{min}) for
the superpotential (\ref{w}) are in correspondance with the five
singlets, $\psi_1,\, \bar{\psi_1},\, A_1,\, A_{24},$ and $S_0$.  The
vanishing of the $D-$terms requires $|\psi_1 | = |\bar{\psi_1}| ,$ and
since $W$ is symmetric under $\psi_1 \leftrightarrow \bar{\psi_1} ,$
the number of relevant equations and singlets is reduced to four.
According to the previous general discussion, one finds the following
solutions:
\begin{itemize}
\item[(i)] a generic vacuum with \GW\  symmetry and components along the 
five directions in proportions that are fixed by the couplings in $W$;
\item[(ii)] a \GG\  symmetric vacuum with vanishing components along
$A_{24}$ and $S_0$ -- there are two non trivial equations for the
components  $|\psi_1 | = |\bar{\psi_1}| $ and $A_1 ;$
\item[(iii)] a $U(3)\otimes SO(4) $ solution with $\psi_1 =
\bar{\psi_1} =0 ,$ $ A_1 / A_{24} = \sqrt{2/3} ,$ as $S_0$ is invariant
under $SO(6)\otimes SO(4)\supset U(3)\otimes SO(4)$ (two non-trivial
equations);
\item[(iv)] a $SO(6)\otimes U(2)$ solution with $\psi_1 = \bar{\psi_1}
=0 ,$ $ A_1 / A_{24} = -\sqrt{3/2} ,$ as $S_0$ is invariant under
$SO(6)\otimes SO(4)\supset SO(6)\otimes U(2)$ (two non-trivial
equations).
\end{itemize}
Even if our choice of chiral multiplets is motivated by the physical
requirement of the breaking of \GUT\  into \GW , there are also extrema
of $W$ where the residual invariance does not contain \GW : \eg , a
$SO(7)$ invariance that corresponds to the little group of a critical
orbit of \irr{16} \op \irr{16*} and is associated to the \irr{45} and
the \irr{54} along their critical orbits with a $SO(8)\otimes SO(2)$
symmetry.

In the cases (ii)--(iv), there is at most one singlet of the residual
symmetry in each \GUT\  irreducible representation which correspond to
a critical orbit. The total number of singlets and equations is reduced
to two as a result of the increased symmetry.

An approximate intermediate symmetry $G_I =$ \flip\ can be obtained by
deforming the vacuum by a fine-tuning on the superpotential couplings,
$(\sqrt{15} \kappa M + 2\lambda \mu) \rightarrow 0 ,$ so that $A_{24}
\rightarrow 2\sqrt{6}A_1 ,$ which gives the \flip\ singlet in the
\irr{45} .  The parameters in $W$ are chosen such that $A'_{24}$ takes
the largest \vev, yielding an intermediate \flip\ symmetry.  The usual
Georgi-Glashow \GG\  is defined by the $\psi_1$ \vev\ which breaks
$(B-L).$ The exact symmetry, once all \vev 's are taken into account,
is \GW .

The implementation of an intermediate \flip\ symmetry requires, besides
the obvious hierarchy in the parameters to define an intermediate
scale, a tuning of the couplings in the model. This approximate
symmetry solution does not have an exact symmetry solution counterpart,
because in the $S_0 \rightarrow 0,\  \bar{\psi_1}\psi_1 \rightarrow 0 $
limit even the \irr{45} component vanishes: This is related to the
absence of a cubic invariant for $A. $

Instead, the Georgi-Glashow \GG\  supersymmetric vacuum is generically
present, without any tuning of the parameters. If one introduces some
hierarchy in the couplings in $W$, namely $ M \gg \mu$ and $h \gg
\lambda ,$ another vacuum possesses an approximate \GG\ symmetry as
$A_1 > S_0 > A_{24} .$ There could be some tunneling between these two
vacua when they get close.  We may conclude that the \flip\ is a less
natural intermediate symmetry unless the required fine tuning is
provided by some mechanism, \eg , the existence of a fixed point.

Although the \irr{45} has a \flip\ invariant solution (in the same
orbit as the Georgi-Glashow solution), there is no solution with that
symmetry since the only non trivial invariants with $A$ in $W$ are
linear in $\psi_1$ and $\bar{\psi_1}$ or $S$ which have no
\flip\ invariant direction. Indeed the direction of $A$ in the $(A_1 ,
\, A_{24} )$ space is settled by a compromise between the alignment to
$\psi_1$ and $\bar{\psi_1}$ to give the Georgi-Glashow \GG\  and the
alignment to $S_0$ along either the $U(3)\otimes SO(4)$ or the
$SO(6)\otimes U(2)$ directions. Therefore, only a particular tuning
brings the \irr{45} along the \flip\ direction. The inclusion of higher
degree polynomial invariants does not prevents  the need for a tuning
in the parameters of the superpotential, which gets more involved.
However, the presence of a quartic non trivial  invariant, $\tr A^4 ,$
allows for a solution with exact \flip . Nevertheless, for  $|\psi_1 |
= |\bar{\psi_1}| \neq 0,$ the \irr{45} chooses the \GG invariant
direction, $A_1 .$

The physical interest of an approximate flipped symmetry seems a
motivation to include chiral multiplets transforming in the \irr{210}
(namely, an antisymetric tensor of $\rank = 4$), $\Phi ,$ since a cubic
\GUT invariant $\Phi ^3$ exists which does not vanish along the
\flip\ invariant direction of $\Phi .$ There are three \GW\ invariant
directions in the \irr{210}, $\Phi_1$, $\Phi_{24}$ and $\Phi_{75} ,$
transforming  as a singlet, a \irr{24} and a \irr{75} under \GG ,
respectively.

The following linear combinations:
\be
\Phi '_{1} = -\frac{1}{5}\Phi_{1}+\frac{2}{5}\Phi_{24}+
\frac{2}{\sqrt{5}}\Phi_{75} \nonumber \\
\Phi '_{24} = \frac{2}{5}\Phi_{1}+\frac{13}{15}\Phi_{24}-
\frac{2}{3\sqrt{5}}\Phi_{75} \nonumber \\
\Phi '_{75} = \frac{2}{\sqrt{5}}\Phi_{1}-\frac{2}{3\sqrt{5}}\Phi_{24}+
\frac{1}{3}\Phi_{75}  \label{75flip}
\ee
transform as \irr{1}, \irr{24}, \irr{75} under \flip , respectively.

Let us first concentrate on the \irr{16} \op \irr{16*} \op \irr{210}
chiral multiplets and the generic superpotential 
\be
\tilde{W} =  m \bar{\psi}\psi + \tilde{M} \Phi ^2 + 
g \bar{\psi} \Phi\psi + \kappa\Phi^3 \label{tildeW}
\ee 
In Table 1, the cubic invariants are explicitly written in terms of
only the \GW\ invariant components of the various \GUT\ multiplets
discussed here, and the corresponding expresions in terms of the
\flip\ relevant directions are displayed in brackets for the first two
invariants and are given by the same expressions with $A,\, \Phi
\rightarrow  A',\, \Phi '$ for the others. The fact that the invariants
containing $\psi_1$ and $\bar{\psi _1}$ couple only to $A_1 ,\, \Phi _1
,$ but to all $A',\, \Phi '$ components disfavours the \flip
\ invariant solution $A'_1$ as we now turn to discuss.

\begin{table} 
\begin{tabular}{|c||c|}
\hline 
&  \\ 
Invariant & Expression limited to \GW\ singlet fields \\ 
& \\  
\hline \hline 
& \\
\irr{16} \ot \irr{16*} \ot \irr{45} & 
$\psi_1\, \bar{\psi _1}\, A_1 \, = \,  \psi_1\, \bar{\psi _1}\,
(\frac{1}{5}A'_1+\frac{2\sqrt{6}}{5} A'_{24}) $ \\  
& \\ 
\hline
& \\
\irr{16} \ot \irr{16*} \ot \irr{210} & 
$\psi_1\, \bar{\psi _1}\, \Phi _1 \, = \,  \psi_1\, \bar{\psi _1}\,
(-\frac{1}{5}\Phi '_{1}+\frac{2}{5}\Phi '_{24}+
\frac{2}{\sqrt{5}}\Phi '_{75} )$ \\  
& \\ 
\hline
& \\
\irr{54} \ot \irr{45} \ot \irr{45} & 
$ S_0\, (\, A^2_{24} - 2\sqrt{6}A_{24}A_1 \,)$ \\  
& \\ 
\hline
& \\
\irr{210} \ot \irr{210} \ot \irr{210} & 
$ \Phi ^3_1 + \frac{7}{27}\Phi ^3_{24} + \frac{8\sqrt{5}}{27}\Phi ^3_{75} 
+ \frac{1}{2}\Phi ^2_{24}\Phi _1 + \frac{5\sqrt{5}}{18}\Phi ^2_{24}\Phi _{75}
- \Phi ^2_{75}\Phi _1 + \frac{8}{9}\Phi ^2_{75} \Phi _{24}$ \\   
& \\ 
\hline
& \\
\irr{210} \ot \irr{45} \ot \irr{45} & 
$ \Phi _1 ( 2A^2_1 - \frac{1}{2}A^2_{24} ) + \Phi _{24}( \frac{1}{3}A^2_{24}  
+ \sqrt{6} A_{24}A_1 ) +  \frac{5\sqrt{5}}{6}\Phi _{75}A^2_{24} $ \\    
& \\ 
\hline
& \\
\irr{45} \ot \irr{210} \ot \irr{210} & 
$ A_1 (3\Phi ^2_1 - 2\Phi ^2_{24} + \Phi ^2_{75}) - {\sqrt{6}}A_{24} 
(\Phi _{24}\Phi _1 + \frac{2}{9}\Phi ^2_{24} - \frac{5\sqrt{5}}{9}
\Phi_{75}\Phi _{24} - \frac{8}{9}\Phi ^2_{75}) $\\
& \\
\hline
& \\
\irr{54} \ot \irr{210} \ot \irr{210} & 
$ S_0 ( - \Phi ^2_{24} - 8\Phi ^2_{75} + 18\Phi _{24}\Phi _1 
+ \frac{10\sqrt{5}}{9}\Phi _{75}\Phi _{24} )  $\\
& \\ 
\hline

\end{tabular}
\end{table}

The presence in Eq.(\ref{tildeW}) of both the quadratic and cubic
invariants for the \irr{210} representation implies the existence of a
vacuum such that $\psi = \bar{\psi} =0$ and $\Phi$ belongs to any
critical  orbit ( excepting those like the one with $SO(6)\otimes
SO(4)$ symmetry for which the cubic invariant vanishes ) including the
one that contains both the \GG\  x U(1) and the \flip\  symmetric
vacua. However, for $|\psi_1 | = |\bar{\psi_1}| \neq 0,$ the \irr{210}
\GG $\otimes U(1)$ invariant \vev\ aligns with the \GG\  invariance of
the $\psi_1 .$ Including a \irr{54} chiral multiplet, the vanishing of
the gradient of $W$ along $\Phi_{75}$ still implies
$g\psi_1\bar{\psi_1} =0.$ Indeed, we cannot put $g=0$ because this
coupling  is the only one that links the  \irr{210} and the \irr{16}
\op \irr{16*}  directions.

Finally, with a \irr{16} \op \irr{16*} \op \irr{45} \op \irr{210}
chiral multiplet, if $\psi_1\bar{\psi_1} \neq 0$ both the  \irr{45} and
the \irr{210} must align to the \GG\ invariant direction. To enforce
the \irr{45} or \irr{210} \vev\ to be along the \flip\ invariant
direction one needs one (two, resp.) tuning conditions to reduce the
number of independent equations.

The necessity of tuning conditions to get supersymmetric vacua
symmetric under \GW\  with a dominant \vev\ along the \flip\  invariant
direction of an irreducible representation is related to the presence
of other singlets of \GW\ in the same representation. This gives rise
to more minimum equations than variables.This problem does not arise
for the Georgi-Glashow \GG\ since there are no other \GW\ singlets in
the \irr{16} (or in the \irr{126}) representation. This is not a reason
not to consider supersymmetric SO(10) models with intermediate flipped
symmetry, as far as the necessary tuning of parameters can be
justified.

More in general, one can easily build supersymmetric patterns of
symmetry breaking to \GW\ with particular intermediate symmetries, when
the \vev\ possessing that invariance belongs to a representation
without other \GW\ singlets. Otherwise, in the most favourable case, a
tuning condition is required.

Finally, we have not considered the \irr{126} \op \irr{126*}
representation since it gives $\Delta _s b < 0$ in the intermediate
\flip\ regime, opposite to what is needed to reach \GUT\ unification.
The \irr{16} \op \irr{16*} \vev\ which gives the scale $M_R$, has
$\|\Delta (B-L) \| =1 $ so that the $\|\Delta L \| =2$ right-handed
neutrino Majorana mass matrix $M_N ,$ must be proportional to $M^2_R .$
If appropriate dimension 5 operators could be generated when the fields
with masses $O( M_U )$ are integrated out, then one would expect $M_N
\sim O( M_R^2 / M_U ) = O(r^2)\times M_U .$ However, at least in the
minimal scheme discussed here, when one adds to the \GUT\ breaking set
of fields the three generations of matter fields in three \irr{16}'s,
these dimension 5 operators are not yielded. Indeed, the \irr{45} or
\irr{210} that couple to the \irr{16} \op \irr{16*} cannot be coupled
to the matter \irr{16}'s without spoiling the \GW\ invariant solution.
This is naturally enforced by requiring the R-parity symmetry which is
anyway needed at the MSSM scale. This forbids contributions $O( M_R^2 /
M_U )$ at the tree-level. Supersymmetry non-renormalization theorems
forbid quantum-loop diagrams to generate them {\it \`a la} Witten
\cite{witt} up to corrections proportional to supersymmetry breaking
soft masses and the R-parity symmetry prevents any mixing through
wave-function renormalization.

We are assuming that the cut-off scale of the \GUT\ theory is $M_{\Pl }
,$ and $M_N $ must originate from non-renormalizable \GUT\ invariant
operators, hence $M_N \sim O( M_R^2 / M_{\Pl } ) = O(r^2)\otimes
10^{14} {\rm GeV}$.  This is consistent with the seesaw mechanism if
the heavier neutrino masses are ${\cal O}(\sqrt{m^2_{\mbox{atm}}}$.
Interestingly enough, this scale is quite close to the one obtained in
(non-supersymmetric) \GUT\ with intermediate Pati-Salam symmetry
\cite{miele}, even if the scale of \BL\ breaking are very different.
This difference might be relevant for baryogenesis through
leptogenesis.  Instead, with intermediate \GG\ such that $M_R$ is above
the unification scale, the right-handed neutrinos are expected to be
heavier, especially with  a \irr{126} \op \irr{126*} breaking of \BL .
In supersymmetric theories these different patterns for right-handed
neutrino masses give rise to different predictions for charged lepton
flavour violating decays.

In contrast with the relatively minimal field content and the use of
low dimension couplings adopted here, many papers have suggested to
achieve the symmetry breaking by selecting only a few suitable
couplings \cite{moha, other} through \adhoc\ discrete symmetries. The absence
of couplings that are gauge invariant in the superpotential usually
leads to new solutions and even flat directions \cite{bdfs} that should
be carefully examined (just as the use of R-parity constraints brings
about the colour breaking vacua in the supersymmetric extensions of the
Standard Model).

\vskip 2truecm

\end{document}